\documentclass[twocolumn,showpacs,preprintnumbers,amsmath,amssymb]{revtex4}
\usepackage{graphicx}
\usepackage{dcolumn}
\usepackage{bm}
\usepackage{color}
\begin{document}

\title{Probing confinement by direct photons and dileptons}

\author{V.V.~Goloviznin$^{1}$, A.V.~Nikolskii$^{2}$, A.M.~Snigirev$^{3,2}$, and G.M.~Zinovjev$^{1}$}

\affiliation{$^{1}$ Bogolyubov Institute for Theoretical Physics, National Academy of Sciences of Ukraine, Kiev 03143, Ukraine}

\affiliation{$^{2}$ Bogoliubov Laboratory of Theoretical Physics, JINR, 141980, Dubna, Russia }

\affiliation{$^{3}$ Skobeltsyn Institute of Nuclear Physics, Lomonosov Moscow State University, 119991, Moscow, Russia }

\date{\today}
\begin{abstract}
The intensive synchrotron radiation resulting from quarks interacting with the collective confining color field in relativistic heavy ion collisions is discussed. The spectrum of photons with large transverse momentum is calculated and compared with the experimental data to demonstrate the feasibility of this type of radiation.
A study of the earlier predicted azimuthal anisotropy in the angular
distribution of dileptons with respect to the three-momentum of the pair is performed as well. This boundary-induced mechanism of lepton pair production is shown to possess the features that are distinctly different from the standard mechanisms and can potentially provide an efficient probe of quark-gluon plasma formation.

\end{abstract}


\maketitle

\section{Introduction}
Removing the mystery flair from the widely spread belief that quark-gluon plasma (QGP) has already been created in relativistic heavy ion collisions at RHIC and LHC one finds enough questions that need to be answered to strengthen this belief and all the more make it a reality. Amongst those the physics of confinement/deconfinement transformation and its experimentally observable signatures remains a key issue~\cite{schukraft}. Since photons and leptons freely leave the plasma medium without interacting with the interior staff, they all are still the most suitable source of direct information on the state of hot hadronic matter (see, e.g., the surveys~\cite{Shen:2016odt,Paquet:2016pnt,Masson:2018}). 

In this context a clear excess of direct photons above the expectation from scaled $pp$ collisions recently observed by the PHENIX~\cite{phenix2010} and ALICE~\cite{alice} Collaborations turned out rather underestimated by the theoretical calculations of thermal QGP radiation. Moreover, unexpectedly these photons carry also the large azimuthal momentum anisotropy~\cite{phenix2011} very similar to the hadronic one. Up to now theoretical calculations still underestimate these challenging measurements~\cite{Shen:2016odt,Paquet:2016pnt}. This tension known as ``direct photon flow puzzle'' remains rather intriguing and exciting although several phenomenological suggestions (see, for example,  ~\cite{Shen:2016odt,Paquet:2016pnt,Chatterjee:2005de,pisarski,pisarski2,pisarski3,chiu,chatterjee,hees,kharzeev,bzdak,liu,linnyk,Zakharov:2016mmc,Zakharov:2017cul} and Refs. therein) have been done to understand an origin of this interesting observation. Fortunately, the situation with improving a description of global photon data overall is gradually becoming more controlled~\cite{vujanovic} despite an increasing number of possible photon sources. In our previous papers~\cite{Goloviznin:2012dy,Goloviznin:2017ggj} we have suggested an alternative mechanism that contributes to the observed anisotropy of direct photons but was yet apparently not taken into consideration in the available phenomenological estimates. The reference is to a ``magnetic bremsstrahlung-like radiation'' (or synchrotron radiation in present terminology) of quarks in the collective color field ensuring confinement. We have found that the total intensity of such a radiation for the hot medium of size 1-10 fm that is expected to be created in ultra-relativistic collisions of heavy ions is comparable with the radiation intensity of standard mechanism of photon and dilepton production in the temperature range of $T= 200-500$ MeV. A relative effect is quantitatively regulated by the three basic parameters: the QGP temperature $T$, the characteristic medium (QGP) size $R$, and the confining force $\sigma$ which are fixed. Possible uncertainties come mainly from the simple modeling of confinement and simplification of the QGP geometry. However, we demonstrate it allows us to obtain the fast estimates in the transparent analytical form.
In this paper we focus on the spectrum of the hard enough bremsstrahlung photons which can be analytically estimated and therefore can be directly compared with experimental data. Our note is organized as follows. First, we describe our theoretical approach and estimate the spectrum of photons with the large enough transverse momenta. Then after comparison with the  data we discuss the peculiarity of angular photon and dilepton distributions which can help us to make a choice of mechanism among other sources of direct photon and dilepton production to probe the confinement features. Finally, in conclusion we summarize the main results. 

\section{Theoretical framework}

Since we develop a pretty original scenario of photon and dilepton production here we start from the basic statements and equations following our last papers~\cite{gol1,gol2,gol3,gol4} to be clear and easy verifiable. We consider the interaction of outgoing color object and intersecting the QGP volume at its boundary come effectively to the appearance of a constant restoring force  $\sigma$. Apparently, this force is acting along the normal to the plasma surface and, as a result, any light quark (anti-quark) at the boundary of system volume moves along a curve trajectory and (as any classical charge undergoes an acceleration) is emitting photons. This magnetic bremsstrahlung radiation is sufficiently intensive for light quarks due to the large magnitude of the ``string tension'' $\sigma \simeq 0.2$~Gev$^2$ as it is fixed, for example, in the chromoelectric flux tube model~\cite{tube1,tube2,tube3}.

A large value of the confining force $\sigma$ results in the large magnitude of characteristic parameter 
$$\chi = ((3/2) \sigma E/m^3)^{1/3}$$ 
for $u$ and $d$ quarks (the strong-field case), where $E$ and $m$ are the energy and mass of the emitting particle, respectively. According to Refs.~\cite{gol1,gol2,gol3,book1,book2} in such a strong field regime the probability of photon emission is independent of the mass of emitting particle and the spectral distribution can be represented as
\begin{equation}
\label{c2}
\frac{dN_{\gamma}}{d\omega dt} = 0.52 e_q^2 \alpha \omega^{-2/3} (\sigma \sin
\varphi/E)^{2/3},~~~~~0 < \omega < E,
\end{equation}
where $\alpha=1/137$ is the fine structure constant, $e_q$ is the quark charge in units of electron charge and $\varphi$ is the angle between the quark velocity and the direction of quark confining force (the normal to the QGP surface in our case). This expression is valid for all frequencies $\omega$ excepting those near $E$.

The time interval of quark motion (the ``radiation time'') in the field $\sigma$ is of order $E/\sigma$. Accurate calculations tracing the dynamics of quark and chromoelectric tube motion give practically the same magnitude~\cite{tube2,tube3}. If the confining force acts along the $z$-axis we have the equation of motion for a quark crossing a surface of QGP volume in the following form:
\begin{eqnarray}
\label{eq-mot}
p_z = \sigma t, p_y = p_{y0}, p_x = p_{x0}, -p_{z0}/\sigma \leq t \leq
p_{z0}/\sigma,
\end{eqnarray}
where $p_{z0}>0, p_{y0}, p_{x0} $ are the initial values of the corresponding components of quark momentum. To move further we suppose that at any instant of time the direction of the emitted photons coincides with the direction of the quark velocity (since an ultra-relativistic particle emits photons at small ($\sim m/E$) angles around the instantaneous direction of the velocity). Then from Eq.~(\ref{c2}) we find the following spectral angular distribution of photons radiated in one quark ``reflection'' from the QGP surface:
\begin{eqnarray}
\label{c3}
\frac{dN_{\gamma}}{d\omega d\Omega}
& = &\int\limits_{-p_{z0}/\sigma}^{p_{z0}/\sigma} dt ~\delta
({\bf n}-{\bf v}(t)) \theta [\omega < p(t)] \nonumber\\
& & \times \frac{0.52 e_q^2 \alpha
\sigma^{2/3}}{\omega^{2/3}}
\frac{\sin^{2/3} \varphi (t)}{p^{2/3} (t)},
\end{eqnarray}
where
${\bf v}(t)$ is the quark velocity vector, 
${\bf n}$ is the unit vector along the photon momentum and
\begin{eqnarray}
p(t) = (p_x^2 + p_y^2 + p_z^2)^{1/2},~~\sin \varphi (t) = (p_x^2 +
p_y^2)^{1/2}/p(t).\nonumber
\end{eqnarray}
Now having the motion laws it is not a great deal to determine the spectral angular distribution of photons radiated per unit time per unit surface area of plasma. Folding~(\ref{c3}) with the flux of quark reaching the boundary and integrating over initial quark momenta we have
\begin{eqnarray}
\label{c4}
\frac{dN_{\gamma}}{dSdt\omega^2 d\omega d\Omega} & = & \frac{1.04g\langle e_q^2
\rangle \alpha}{(2\pi)^3 \sigma^{1/3}}~~ \frac{3}{7} \omega^{2/3} \sin^{2/3}
\varphi_0 \nonumber \\
& & \times \int\limits_1^{\infty} d \xi \exp\Bigg(-\frac{\omega}{T} \xi\Bigg) (\xi^{7/3}-1),
\end{eqnarray}
where  $\langle e_q^2\rangle=e^2_u+e^2_d$,~~~
$e_u$ and  $ e_d$ are the $u$- and $ d$-quark charges, 
$g =$ spin $\times$ color = 6 is the number of quark degrees of freedom, 
$T$ is the plasma temperature.
$\varphi_0$ is the angle between the normal to QGP surface and the direction of emitted photons. The total number of radiated photons can be obtained from Eq.~(\ref{c4}), integrating over $d\omega$ and $d\Omega$ in the following form
\begin{eqnarray}
\label{c5}
dN_{\gamma} /dSdt = A \langle e_q^2 \rangle \alpha T^{11/3} \sigma^{-1/3},
\end{eqnarray}
where
$A = 3.12g\cdot 2^{5/3} \Gamma^2 (4/3) /(2\pi)^2 \simeq 1.2$, $\Gamma$ is the gamma function.

The equations~(\ref{c2}) and~(\ref{c4}) are the key results in the theory of synchrotron radiation for the QGP case. We would like to stress once more an existence of the boundary bremsstrahlung is based on three quite realistic assumptions: 1) confinement; 2) the presence of relativistic light quarks ($u$ and $d$ quarks) in the hot medium; 3) the semi-classical nature of their motion. This quasi-classical treatment is obviously grounded on the fact that the de Brogle wave lengths of medium constituents are perfectly small being compared with the size of their localization region (QGP). 

In order to get fast phenomenological estimates one needs a specification of the QGP geometry and evolution. In this point we follow by a ``well-trodden''  simplified way. In the picture of employing a hydrodynamical scaling solution~\cite{bjorken}, one has a cylindrically symmetric plasma volume (for central collisions) expanding in the longitudinal directions. Taking for the QGP an ideal gas equations of state, we have
\begin{equation}
\label{c8}
T = T_0(\tau_0/\tau)^{1/3},
\end{equation}
where $T_0$ is the temperature at the proper time $\tau_0$ of hydrodynamic stage.

To exclude the uncertainties in these initial parameters $T_0$ and $\tau_0$ we have compared our mechanism of photon radiation with the standard photon radiation from the QGP volume (``Compton scattering of gluons'', $gq \rightarrow \gamma q$ and annihilation of quark-antiquark pairs, $q{\bar q} \rightarrow \gamma g$) dealing with the approximate analytical estimates~\cite{sinha, sinha2}. At such a normalization we have found~\cite{Goloviznin:2012dy,Goloviznin:2017ggj,gol1,gol2,gol3,gol4} that the functional distinction between the proposed mechanism and the ``standard'' volumetric one is mainly determined by the parameter that is just the dimensionless combination as
 \begin{equation}
\label{ratio}
 (rT^{1/3}_c\sigma^{1/3})^{-1}. 
\end{equation}
One should note that it is $\simeq 1$ on setting $r\simeq 0.6$ fm at $T_c=$ 200 MeV, here $r$ is the cylinder radius, $T_c$ is the phase transition temperature (not the initial $T_0$ !). Thus, we have concluded~\cite{Goloviznin:2012dy,Goloviznin:2017ggj,gol1,gol2,gol3,gol4} that the quark interaction with the collective confining color field  results in an intensive enough radiation of the magnetic bremsstrahlung type (synchrotron radiation) to be observed in principle at least at the level of total rate.

Now we are well armed to present a more convincing evidence in favor of the boundary bremsstrahlung appealing to experimental data directly.

\section{The spectrum of photons with large transverse momenta versus experimental data}

Clearly, the dominant contribution to the total photon number (Eq.~(\ref{c5}) comes from soft $\gamma$-quanta, but the presence of large background of $\pi^{0} \rightarrow 2\gamma$ decays and other numerous sources makes practically impossible to employ it as an indicator of plasma formation events. Then a task to extract the spectrum of $\gamma$-quanta with large transverse momenta becomes fun and practical. Fortunately, our approach allows us to make a proper estimate.  We obtain from our ``master equation'' Eq.~(\ref{c4})
\begin{eqnarray}
\label{c17}
\frac{dN_{\gamma}}{dSdtk_{\perp}dk_{\perp}} & = &\frac{1.04g\langle e_q^2 \rangle
\alpha k_{\perp}^{5/3}}{(2\pi)^3 \sigma^{1/3}} \frac{6}{7} \int
\limits_0^{2\pi} d \alpha \int\limits_0^{\infty} dx 
\int \limits_1^{\infty} d \xi \nonumber\\ 
& & \times (x^2+\sin^2 \alpha)^{1/3} (\xi^{7/3}-1)\nonumber\\
& & \times \exp\Bigg[-\frac{k_{\perp}(1+x^2)^{1/2}}{T}
\xi \Bigg].
\end{eqnarray} 
This  Eq.~(\ref{c17}) can be considerably simplified in the limit $k_{\perp} \gg T$ and the spectrum can be written as
\begin{equation}
\label{c18}
\frac{dN_{\gamma}}{dSdtk_{\perp}dk_{\perp}} = C\langle e_q^2 \rangle \alpha
\sigma^{-1/3} k_{\perp}^{-5/6} T^{5/2} e^{-k_{\perp}/T}, 
\end{equation}
where $C=0.52g\cdot 2^{1/6}\Gamma^2(5/6)\pi^{-5/2}/\Gamma(5/3) \simeq 0.29$ is the result of averaging over angles.
 
New spectrum (\ref{c18}) is already quite suitable for analysis to be compared with the experimental data after the appropriate specification of the QGP geometry and evolution. Then for central collisions we have a cylindrically symmetric plasma volume expanding in the longitudinal directions in accordance with Eq.~(\ref{c8}). Therefore integrating  
Eq.~(\ref{c18}) over the QGP surface takes into account an evolution and gives
\begin{eqnarray}
\label{c19} 
\frac{d^2N_{\gamma}}{2\pi k_{\perp }dk_{\perp}dy} = \int \frac{dN_{\gamma}}{dSdt k_{\perp} dk_{\perp}} r
\tau d\tau = C \langle e_q^2 \rangle \alpha 3(\tau_0T_0^3)^2 r  \nonumber \\
\times   k_{\perp}^{-13/3} \sigma^{-1/3}
\Bigg[\Gamma\Bigg(\frac{7}{2};\frac{k_{\perp}}{T_0}\Bigg)
-\Gamma\Bigg(\frac{7}{2};\frac{k_{\perp}}{T_c}\Bigg)\Bigg]~~
\end{eqnarray}
with
\begin{eqnarray}
\Gamma(n,\alpha_1)-\Gamma(n,\alpha_2)=\int\limits_{\alpha_1}^{\alpha_2}
dtt^{n-1} e^{-t} \nonumber
\end{eqnarray}
and $y$ is the rapidity.

\begin{figure}
\label{fig1}
\includegraphics[width=9.00cm]{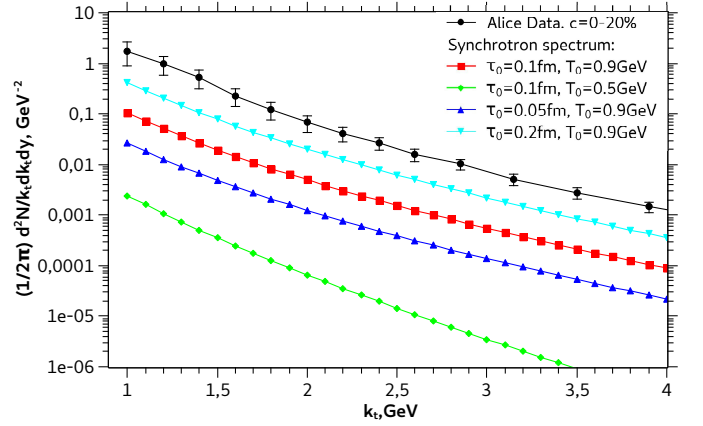}
\caption{Spectra of synchrotron photons for different $T_0$ and $\tau_0$. $T_0$ is the temperature at the moment of proper time $\tau_0$, when the hydrodynamic regime starts. Experimental results taken from~\cite{alice} for the centrality 0-20 $\%$ at a center-of-mass energy per nucleon pair ($\sqrt{ s_{NN}}$) of 2.76 TeV.}
\end{figure}

\begin{figure}
\label{fig2}
\includegraphics[width=9.00cm]{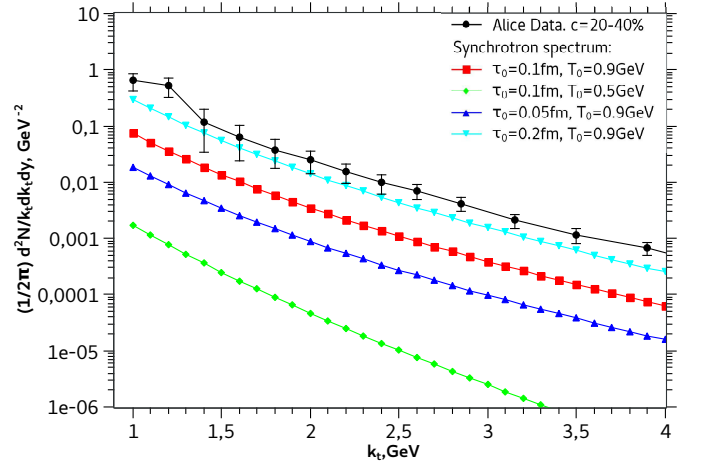}
\caption{Spectra of synchrotron photons for different $T_0$ and $\tau_0$. Experimental results taken from~\cite{alice} for the centrality 20-40 $\%$ at a center-of-mass energy per nucleon pair ($\sqrt{ s_{NN}}$) of 2.76 TeV.}
\end{figure}

In Fig.~1 this spectrum (\ref{c19}) is presented for the different input parameters  $T_0$ and $\tau_0$ at the transverse size of QGP system fixed as $r=10$ fm ($T_c=$~0.2 GeV and $\sigma = 0.2$~Gev$^2$ were fixed earlier). Experimental data from Ref.~\cite{alice} in the most central Pb-Pb collisions for the 0-20 $\%$ centrality class at mid-rapidity $y$  in the transverse momentum range $1 < k_t < 4$ GeV/$c$ are also given for comparing. The absolute normalization of the synchrotron radiation is very sensitive to the input parameters  $T_0$ and $\tau_0$ which are, however, an attribute of any hydrodynamics inspired model, but not the feature of the boundary bremsstrahlung itself. That is why we have early preferred to compare the total intensity with that of the volume mechanism of photon production, where the uncertainties above is not significant at all. We emphasize once more the scenario developed here has no special additional free parameters besides of those used in all hydrodynamics inspired models.

At the ``nominal'' values of these input parameters $T_0=0.9$~GeV and $\tau_0=0.1$~fm (used often for Pb-Pb collisions at the LHC energies, see, for instance, Ref.~\cite{Lokhtin:2014vda}) the synchrotron radiation contributes to the experimentally measured rate of direct photons at the level of $10\%$. In Fig.~2 we compare the spectrum (\ref{c19}) with experimental data for the 20-40 $\%$ centrality class calculating also the “characteristic” transverse size of QGP system over the simple scaling option~\cite{Lokhtin:2005px}: $r(b)=r\sqrt{1-b/2r}$ with $b$ being the impact parameter and $c=(b/2r)^2$ being the geometrical centrality. For this 20-40 $\%$ centrality class $<r(b)> \simeq 7$ fm is used and a constant $C$ of averaging in Eq.~(\ref{c18}) is not recalculated for simplicity since its variation is expected to be unessential.  The relative weight of the boundary bremsstrahlung should be larger for the larger centrality classes as it was revealed in our previous works~\cite{Goloviznin:2012dy,Goloviznin:2017ggj,gol1,gol2,gol3,gol4}: a factor $(rT^{1/3}_c\sigma^{1/3})^{-1}$  for the total intensity and a factor $(rk_{\perp}^{1/3}\sigma^{1/3})^{-1}$ for the photon rate with the large transverse momenta. Figure~2 is just presented only to illustrate the growth of the relative weight of the boundary bremsstrahlung with the decrease of the characteristic size of QGP system (the increase of the impact factor) being the same with other parameters changing.

The uniform transverse temperature profile with a sharp edge is implicitly assumed in the above estimates. In fact, the initial energy deposition is maximum in the middle of highly excited system decreasing towards the periphery (boundary) in the transverse direction. To take into account this finite temperature gradient in the transverse direction now we will assume that the temperature is equal $T_0 (\tau_0 /\tau)^{1/3}$ at the cylindrical axis as before, but the effective boundary temperature $T_s$, at which the photons are emitted, is less and close to the confining temperature $T_c$ being the constant during the plasma evolution. In this case we have the following analytical spectrum which should be compared with experimental data for the 0-20 $\%$ centrality class
\begin{eqnarray}
\label{c19-s} 
\frac{d^2N_{\gamma}}{2\pi k_{\perp }dk_{\perp}dy}& = &\int \frac{dN_{\gamma}}{dSdt k_{\perp} dk_{\perp}} r
\tau d\tau = C \langle e_q^2 \rangle \alpha \frac {\tau_f ^2-\tau_0^2}{2} \nonumber \\
& & \times r  k_{\perp}^{-5/6} T_s^{5/2} \sigma^{-1/3} \exp{[-k_{\perp}/T_s]}
\end{eqnarray}
with the total QGP ``life time'' $\tau_f= (T_0/T_c)^3 \tau_0$ estimated as for the uniform temperature transverse profile.

In Fig.~3 this spectrum (\ref{c19-s}) is presented for the different
boundary temperature $T_s$ =  $T_c$= 0.2~GeV,  $T_s$= 0.3~Gev, $T_s$ = 0.4~GeV versus ALICE data at $T_0$= 0.9~GeV,  $\tau_0$= 0.1~fm (other parameters are the same as in Fig.~1). One can see that at the reasonable effective boundary temperature $T_s$= 0.3~GeV the synchrotron radiation contributes to the experimentally measured rate of direct photons meaningfully (at the level of 5~$\%$) practically in the all transverse momentum range $1 < k_t < 4$~GeV/$c$. At the most pessimistic conservative effective boundary temperature $T_s$ = $T_c$= 0.2~GeV the contribution under consideration is notable in the transverse momentum range $ k_t \simeq 1$~GeV/$c$ only. But in this last case the effect estimate is, in fact, based on a minimum.

\begin{figure}
\label{fig3}
\includegraphics[width=9.00cm]{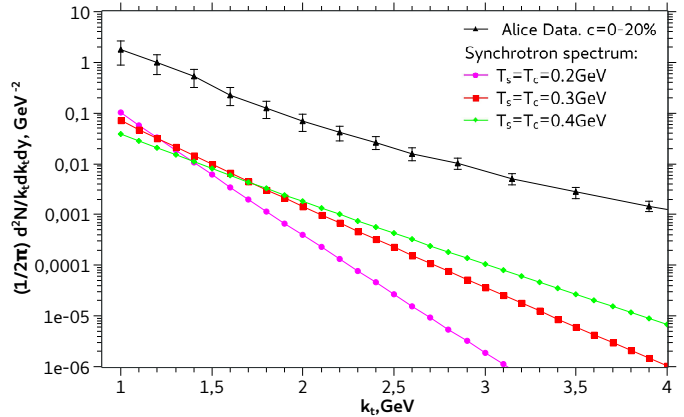}
\caption{Spectra of synchrotron photons for different $T_s$. Experimental results taken from~\cite{alice} for the centrality 0-20 $\%$ at a center-of-mass energy per nucleon pair ($\sqrt{ s_{NN}}$) of 2.76 TeV.}
\end{figure}

Thus, the synchrotron radiation can really help in improving a description of global photon data (the upper curves in Fig.~1 and Fig.~2 with the input parameters $T_0=0.9$~GeV and $\tau_0=0.2$~fm is very impressive). Indeed, in order to draw a more definite conclusion, further investigations should be done and they should include, in particular, a proper comparison with other sources of photons and the detailed elaboration of QGP geometry and its evolution. However, in any case now we have a good ``experimental'' foundation in favor of such an original mechanism. One should note that the availability of other numerous sources does not allow us to select the contribution of the boundary bremsstrahlung unambiguously in spite of the different dependence on the temperature, transverse momenta and system size  (which is not so crucial to distinguish among other sources). In the next Section we discuss the possibility to select this original source of photon and dilepton production among the other known sources basing on its unique feature --- a high degree of photon polarization.

\section{The peculiarities of angular distributions and discussion}

The synchrotron radiation is characterized by a high degree of photon polarization and the phenomena of electron self-polarization~\cite{book2} that clearly distinguishes this mechanism from the ``standard'' mechanism of photon production in the Compton scattering and annihilation. When the photon polarization is taken into account, the equations corresponding to~(\ref{c2}) have the form
\begin{equation}
\label{c11}
\frac{dN_1}{d\omega dt} = \frac{1}{4}\frac{dN_{\gamma}}{d\omega dt},
~\frac{dN_2}{d\omega dt} = \frac{3}{4}\frac{dN_{\gamma}}{d\omega dt},~
\frac{dN_l}{d\omega dt} = \frac{1}{2}\frac{dN_{\gamma}}{d\omega dt}
\end{equation}
where $\frac{dN_{\gamma}}{d\omega dt}$ is determined by Eq.~(\ref{c2}) and corresponds to the situation where a summation over the photon polarization is performed.
$l=1$ describes a right-handed circularly polarized photon and
$l=-1$ describes a left-handed circularly polarized photon.
$N_1$ corresponds to linear polarization of the photon along the vector ${\bf e}_1$, $N_2$ corresponds to linear photon polarization along the vector ${\bf e}_2$. The vectors ${\bf e}_1$ and ${\bf e}_2$ are given by
\begin{equation}
{\bf e}_1 = \frac{{\bf \sigma}\times {\bf k}}{|{\bf \sigma} \times{\bf k}|},
~~~~~{\bf e}_2 = \frac{{\bf k \times e}_1}{|{\bf k \times e}_1|}, \nonumber
\end{equation}
where ${\bf k}$ is the photon momentum. 

In our approximation the effect of quark self-polarization (as the result of magnetic bremsstrahlung radiation) is absent, because the terms linear in the quark polarization (in the case of polarization along the vector ${\bf e}_1$) arise only in the next order of the expansion in inverse powers of the large parameter $\chi$. We recall that for ultra-relativistic electrons moving in a magnetic field the parameter $\chi$ is usually small and the self-polarization effect does occur~\cite{book2}.

As to the photons it is usual to have a high degree of polarization along the vector ${\bf e}_2$. This vector is perpendicular to the photon momentum and lies in the plane formed by the photon momentum and the normal to QGP surface (the direction of the confining force $\sigma$). Thus, the presence of a photon polarization is closely related to the geometrical feature of the QGP volume over whose surface we should integrate. In collision of relativistic heavy ions there is a special direction, it is just the collision axis. In the approach dealing with a hydrodynamical scaling solution~\cite{bjorken}, one has a cylindrically symmetric plasma volume (for central collisions) expanding in the longitudinal directions and the calculations for the final polarization can be done in the explicit form~\cite{gol2}.

The primary degree of polarization
\begin{equation}
 P = \Bigg(\frac{dN_2}{d\omega dt} - \frac{1}{2}
\frac{dN_{\gamma}}{d\omega dt}\Bigg) \Bigg/ \frac{dN_{\gamma}}{2d\omega dt}=\frac{1}{2} 
\end{equation} 
reduces to about 20$\%$ for a plasma with a cylindrically symmetric volume after the transparent, but laborious calculations~\cite{gol2}.
These photons are dominantly polarized along the normal to the plane 
spanned by the momentum of registered photons and the cylinder axis. The appearance of such a polarization is closely connected to the choice of the confining force direction and its value is virtually insensitive to the parameters regulating an intensity of radiation. Hence, these boundary photons can also be polarized for other ``non-ideal'' shapes of QGP surface possessing this decisive feature. The corresponding calculations of the polarization degree in a lucid form can be very complicated.

Since the difficulties of registering photon polarization entail many problems for experimental search for this effect, we have recently suggested in Ref.~\cite{Goloviznin:2017ggj} (see, also Ref.~\cite{Speranza:2018osi}) that observing lepton-pair spectra resulting from the polarization of intermediate photon could be potentially an efficient probe of QGP~\cite{Baym:2017qxy,Baym1} if formed in collisions of ultra-relativistic ions.

We have found that the lepton distribution in the radiation angle takes the form
\begin{eqnarray}
\label{f4}
\frac{dN}{dtd\Omega_1} & = &\frac{\alpha n}{2\pi k^0} \int \frac{p^2dp}{p_1^0 (k^0 - p_1^0)} \delta[f(p)]\\
& & \times  \Bigg[\frac{k^2+2\mu^2}{3}
- \frac{2}{3} \delta p^2 \sin^2 \theta_1
\cos 2\phi_1
\Bigg]\nonumber
\end{eqnarray}
at the decay of massive photons with the four-momentum $k$ into a lepton pair with the 
four-momenta of the lepton $p_1$  and anti-lepton $p_2$. Deriving Eq.~(\ref{f4}) we define $ n(1+\delta)/3$ as the photon number of states with polarization vector $e_1$, $n(1-\delta)/3$ as the photon number of the states with polarization vector $e_2$ and $ n/3$ as the same with polarization vector $e_3$, and choose the reference frame with the $z$ axis directed along the three-vector ${\bf k}$ and the $x$ and $y$ axes tallying with the directions of ${\bf e_1}$ and ${\bf e_2}$, and
$$e_1=\{0,~1,~0,~0\},~e_2=\{0,~0,~1,~0\},$$
$$e_3=\{|{\bf k}|/\sqrt{k^2},~0,~0,~k^0/\sqrt{k^2}\},
~k=\{k^0,~0,~0,~|{\bf k}|\},$$ 
$$p_1=\{\sqrt{p^2+\mu^2},~ p\sin\theta_1 \cos
\phi_1,~p\sin\theta_1  \sin\phi_1,~p\cos\theta_1\}.$$

In the situation when the photons are unpolarized or have longitudinal polarization (along the vector $e_3$) the angular lepton distribution is independent of the azimuthal angle $\phi_1$ . However, if a massive photon has transverse (in the three-dimensional space) polarization ($\delta$ is not zero) a characteristic dependence on the azimuthal angle $\phi_1$ takes place. In the regime of strong field (the large magnitude of characteristic parameter $\chi$), the intermediate photons could be considered up to the masses $\sqrt{k^2} \simeq \sqrt{\sigma}=0.45$ GeV as having a small virtuality and their properties are quite close to real photons~\cite{zhulego}. It means these photons are transversely polarized with practically the same degree of polarization $\delta$  about 20$\%$ as calculated for real photons at a cylindrically symmetric geometry. Thus, the ``bremsstrahlung'' leptons could be identified by measuring their angle anisotropy that is absent in the Drell-Yan mechanism and the ``standard'' volumetric mechanism. The possibility of observing this effect in experiments is supported by our estimates made here and before for the rate of the boundary photons and dileptons.

In our previous works~\cite{Goloviznin:2012dy,Goloviznin:2017ggj} we have also drawn the attention to the fact that the synchrotron radiation will be non-isotropic for the non-central collisions as another distinctive feature. Indeed, photons are emitted mainly around the direction determined by the normal to the ellipsoid-like surface. In the transverse ($x$-$y$) plane (the beam is running along ($z$)-axis) the direction of this normal (emitted photons) is determined by the spatial azimuthal angle $\phi_s=\tan^{-1}(y/x)$ as
\begin{equation}
\label{phi}
\tan(\phi_{\gamma}) = (r_x/r_y)^2 \tan(\phi_s).
\end{equation}
The shape of quark-gluon system surface in transverse plane is controlled by the radii $r_x= r (1-\epsilon)$ and $r_y = r \sqrt{1-\epsilon^2}$ with the eccentricity $\epsilon =b/2r$ ($b$ is the impact parameter, $r$ is the radius of the colliding (identical) nuclei). In this case the ``mean normal'' is not zero and is equal to
\begin{equation}
\label{normal}
\int_0^{2\pi} d\phi_s \cos(2\phi_{\gamma}) /(2\pi)= \epsilon.
\end{equation}
It means that the photon azimuthal anisotropy characterized by the second Fourier component
\begin{equation}
\label{v2}
v_2^{\gamma} = \frac{\int d\phi_{\gamma}\cos(2\phi_{\gamma})(dN^{\gamma}/d\phi_{\gamma})}{\int d\phi_{\gamma}(dN^{\gamma}/d\phi_{\gamma})},
\end{equation}
is not zero as well and is simply proportional to this ``mean normal''
\begin{equation}
\label{v2cos}
v_2^{\gamma} \propto \epsilon.
\end{equation}
The coefficient of elliptic anisotropy for dilepton pairs (the study  suggested in Ref.~\cite{chatterjee}) will be also proportional to the eccentricity of QGP system as it takes place for the bremsstrahlung real photons and can be experimentally measured.

\section{Conclusion}
The main message of present investigation, as we see it, looks quite transparent. The synchrotron radiation should be almost with necessity taken into consideration at a description of global photon and dilepton data. For the most central collisions (where the relative effect due to the synchrotron radiation is minimal compared to the other volume sources because of the size factor $1/r$) the boundary photons contribute to the experimentally measured rate of direct photons at the level of $10\%$. 

In order to distinguish this inspiring mechanism of radiation unambiguously we suggest to study the noticeable specific anisotropy in the angle distribution of leptons with respect to the three-momentum of pair. The origin of such an anisotropy is seemingly rooted in the existence of characteristic direction in the field where the quarks are moving (what was not discussed in the other phenomenological considerations to the best of our knowledge). Besides, another indicative (and convincing) feature could be considered a non-isotropic character of synchrotron radiation for the non-central collisions, since the photons are dominantly emitted around the direction fixed by a normal to surface, and its non-zero ``mean'' value.

\begin{acknowledgments}
Discussions with A.~Borissov, K.A.~Bugaev, L.V.~Malinina, S.N.~Nedelko, J.-F.~Paquet, D.Yu.~Peresunko, O.V.~Teryaev and V.D. Toneev are gratefully acknowledged.
The paper was supported by the Goal-Oriented Program of Cooperation between CERN and National Academy of Science of Ukraine "Nuclear Matter under Extreme Conditions" (agreement CC/1-2018) and was also partially supported by Russian Foundation for Basic Research (grant 18-02-00155).
\end{acknowledgments}



\begin{thebibliography}{99}
\bibitem{schukraft} J. Schukraft, arXiv:1705.02646 [hep-ex].
\bibitem{Shen:2016odt}C.~ Shen, Nucl. Phys. A {\bf 956}, 184 (2016).
\bibitem{Paquet:2016pnt}J.-F.~Paquet, J. Phys. Conf. Ser. {\bf 832} (1), 012035 (2017).
\bibitem{Masson:2018}E.~Masson, arXiv:1811.02220 [hep-ex].
\bibitem{phenix2010} A.~Adare {\it et al.} (PHENIX Collaboration), Phys. Rev. Lett. {\bf 104}, 132301 (2010).
\bibitem{alice} J.~Adams {\it et al.} (ALICE collaboration), Phys. Lett. B {\bf 754}, 235 (2016).
\bibitem{phenix2011} A.~Adare {\it et al.} (PHENIX Collaboration), Phys. Rev. Lett. {\bf 109}, 122302 (2012).
\bibitem{Chatterjee:2005de} R. Chatterjee,  E. S. Frodermann, U.W. Heinz, and D.K. Srivastava, Phys. Rev. Lett. {\bf 96}, 202302 (2006).
\bibitem{pisarski} C. Gale, Y. Hitaka, S. Jeon, S. Lin, J.-F.~Paquet, R. Pisarski, D. Satow, V. Skokov, and G. Vujanovic, Phys. Rev. Lett. {\bf 114}, 072301 (2015). 
\bibitem{pisarski2}R. Pisarski, Phys. Rev. D {\bf 74}, 121703 (2006).
\bibitem{pisarski3}A.~Dumitru, Y.~Guo, T.~Hidaka, C.P.K.~Altes, and R.~Pisarski, Phys. Rev.D {\bf 86}, 105017 (2012).
\bibitem{chiu} M. Chiu, T.K. Hemmick, V. Khachatryan, A. Leonidov, J. Liao, and L. McLerran, Nucl. Phys. A {\bf 900}, 16 (2013).
\bibitem{chatterjee} R. Chatterjee,  D.K. Srivastava, U.W. Heinz, and C. Gale, Phys. Rev. C {\bf 75}, 054909 (2007).
\bibitem{hees} H. van Hees, C. Gale, and R. Rapp, Phys. Rev. C {\bf 84}, 054906 (2011).
\bibitem{kharzeev} G. Basar, D. Kharzeev, and V. Skokov, Phys. Rev. Lett. {\bf 109}, 202303 (2012).
\bibitem{bzdak} A. Bzdak and V. Skokov, Phys. Rev. Lett. {\bf 110}, 192301 (2013).
\bibitem{liu} F.-M. Liu and S.-X. Liu, Phys. Rev. C {\bf 89}, 034906 (2014).
\bibitem{linnyk} O. Linnyk, V.P. Konchakovski, W. Cassing, and E.L. Bratkovskaya, Phys.Rev. C {\bf 88}, 034904 (2013).
\bibitem{Zakharov:2016mmc}B.G.~Zakharov, Eur. Phys. J. C {\bf 76}, 609 (2016).
\bibitem{Zakharov:2017cul}B.G.~Zakharov, JETP Lett. {\bf 106}, 283 (2017).
\bibitem{vujanovic} G. Vujanovic, J.-F.~Paquet, S. Ryu, C. Shen, G. Denicol, S. Jeon, C. Gale, and U. Heinz, arXiv:1704.04687 [nucl-th], Proc. of QM 2017. 
\bibitem{Goloviznin:2012dy}V.V. Goloviznin, A.M. Snigirev, and G.M. Zinovjev, JETP Lett. {\bf 98}, 61 (2013).
\bibitem{Goloviznin:2017ggj}V.V. Goloviznin, A.M. Snigirev, and G.M. Zinovjev, JETP Lett. {\bf 107}, 527 (2018).
\bibitem{gol1} V.V. Goloviznin, G.M. Zinov'ev, and A.M. Snigirev, Yad Fiz. {\bf 47}, 886 (1988) [Sov. J. Nucl. Phys. {\bf 47}, 561 (1988)].
\bibitem{gol2} V.V. Goloviznin, G.M. Zinov'ev, and A.M. Snigirev, Yad Fiz. {\bf 48}, 1826 (1988) [Sov. J. Nucl. Phys. {\bf 48}, 1099 (1988)].
\bibitem{gol3} V.V. Goloviznin, A.M. Snigirev, and G.M. Zinovjev, Z. Phys. C {\bf 38}, 255 (1988).
\bibitem{gol4} V.V. Goloviznin, A.M. Snigirev, and G.M. Zinovjev, Z. Phys. C {\bf 45}, 335 (1989).
\bibitem{tube1} A. Casher, H. Neuberger, and S. Nussinov, Phys. Rev. D {\bf 20}, 179 (1979). 
\bibitem{tube2} B. Banerjee, N. Glendenning, and T. Matsui, Phys. Lett. B {\bf 127}, 453 (1983).
\bibitem{tube3} V.V. Goloviznin, A.M. Snigirev, and G.M. Zinovjev,  Phys. Lett. B {\bf 211}, 167 (1988).
\bibitem{book1}V.B.~Berestetskii, E.M.~Lifshitz, and L.P.~Pitaevskii, Quantum Electrodynamics, 2nd ed. (Pergamon Press, Oxford, 1982) [ Russ. original, Nauka, Moscow, 1980, vol.IV].
\bibitem{book2}A.A.~Sokolov and I.M.~Ternov, Synchrotron Radiation [in Russian] (Nauka, Moscow, 1966).
\bibitem{bjorken} J.D. Bjorken, Phys. Rev. D {\bf 27}, 140 (1983).
\bibitem{sinha} B. Sinha, Phys. Lett. B {\bf 128}, 91 (1983).
\bibitem{sinha2}
D. Srivastava and B. Sinha, Phys. Rev. C {\bf 64}, 034902 (2001).
\bibitem{Lokhtin:2014vda} I.P.~Lokhtin, A.A. Alkin, and A.M.~Snigirev, Eur. Phys. J. C {\bf 75}, (2015) 452.
\bibitem{Lokhtin:2005px} I.P.~Lokhtin and A.M.~Snigirev, Eur. Phys. J. C {\bf 45}, (2006) 211.
\bibitem{Speranza:2018osi} E.~Speranza, A.~Jaiswal, and B.~Friman, Phys. Lett. B {\bf 782}, 395 (2018).
\bibitem{Baym:2017qxy}G.~Baym, T.~Hatsuda, and M.~Strickland, Phys. Rev. C{\bf 95}, 044907 (2017).
\bibitem{Baym1} G. Baym and T. Hatsuda, PTEP 2015, no.3, 031DO1 (2015).
\bibitem{zhulego} V.G. Zhulego, V.N. Rodionov, and A.I. Studenikin, Yad Fiz. {\bf 36}, 524 (1982) [Sov. J. Nucl. Phys. {\bf 36}, 306 (1982)].
\end{thebibliography}
\end{document}